%% file: root.tex
\documentclass[letterpaper, 10 pt, conference]{ieeeconf}  

\IEEEoverridecommandlockouts                              

\usepackage[utf8]{inputenc}
\usepackage{bigints}
\usepackage{color,soul}
\usepackage{array}
\newcolumntype{P}[1]{>{\centering\arraybackslash}p{#1}}
\newcolumntype{M}[1]{>{\centering\arraybackslash}m{#1}}
\usepackage[thinlines]{easytable}
\usepackage{multirow}
\usepackage{xcolor,colortbl}

\definecolor{Gray}{gray}{0.85}
\definecolor{LightCyan}{rgb}{0.88,1,1}
\newcolumntype{a}{>{\columncolor{Gray}}c}
\usepackage{hyperref}
\usepackage{amsmath}
\usepackage{lipsum}
\usepackage{ams}
\usepackage{multicol, blindtext}

   \usepackage[pdftex]{graphicx}     
   \graphicspath{{../pdf/}{../jpeg/}{./image/}}    
   \DeclareGraphicsExtensions{.pdf,.jpeg,.png,.jpg}

\usepackage{amsmath}
\usepackage{bm}
\graphicspath{{Figures/}}
\usepackage[normalem]{ulem}
\usepackage[backend=biber,
style=ieee,
citestyle=numeric 
]{biblatex}
\graphicspath{ {./image/} }

\title{\LARGE \bf
Low-bandwidth Modular Mathematical Modeling of DC Microgrid Systems for Control Development with Application to Shipboard Power Systems
}

\author{M. M. Bijaieh$^*$, \textit{Member, IEEE}, S. Vedula, and O. M.  Anubi, \textit{Member, IEEE}
\thanks{Authors are with the Department of Electrical and Computer Engineering, Center for Advanced Power systems, Florida State University, Tallahassee, FL 32310, USA
{\tt\small E-mail: \{mmohammadibijaieh,svedula, oanubi\}@fsu.edu}}%

}

\begin{document}

\maketitle
\thispagestyle{empty}
\pagestyle{empty}

\begin{abstract}
In recent years, DC and AC microgrid (MG) systems have attracted a major attention due to various potential for integration of future technology into conventional systems and control. The integration of such technology requires appropriate tools for complex design, analysis and optimization. This paper presents a mathematical low-bandwidth modeling (LBM) approach that can be used for control development in DC and further be extended to AC MG systems. In this work, first a simplified mathematical model of a medium voltage DC (MVDC) shipboard MG system is presented, next, the overall system-level connection convention is presented to display the overall mathematical coupling of the individual sub-systems, then, a simplified example of the control development is presented, and last, the overall system under a test scenario is implemented in Simulink Real-time.

\end{abstract}

\section{INTRODUCTION}\label{Sec: Introduction}
\input{introduction} 

\section{DC MICROGRID CONFIGURATION}\label{Sec: System_Model}
\input{system_model}

\section{DC MICROGRID MATHEMATICAL MODELING: A GENERALIZED EXAMPLE}\label{Sec: dc_MG_model}
\input{dc_MG_model}

\section{CONTROL DEVELOPMENT} \label{Sec: Control_Development}
\input{control_development}

\section{ILLUSTRATIVE EXAMPLE}\label{Sec: Simulation}
\input{simulation}

\section{CONCLUSIONS}\label{Sec: Conclusion}
\input{conclusion}

\section{ACKNOWLEDGEMENT}\label{Sec: ACKNOWLEDGEMENT}
\input{Acknowledgement}







\medskip
\printbibliography

\end{document}

%% file: introduction.tex
The MG concept was first presented in \cite{2001_Lasseter}, where flexible clusters of energy generating, storing, and dissipating sub-systems were envisioned for effective integration of micro-sources. MG configurations are typically classified as islanded, and grid-tied (networked) systems \cite{2001_Lasseter,2007_Hatz_Microgrids}. Considering their control, general approaches are centralized and distributed operations \cite{2011_Gue}. The former is capable of achieving high levels of performance and optimization but may be susceptible to single points of failure and expandability limitations. The latter offers flexibility at the cost of computational complexity and increased control surface which further exposes the system to more uncertainty including cyber attacks. These phenomena, in turn, further increase the complexity of the distributed MG control.

Analogous to the hierarchical nature of control and resource allocation strategies, \cite{2011_Gue} proposed a hierarchical control architecture for AC and DC MGs that includes level zero, primary, secondary and tertiary controllers. Currently, these control levels are widely used to provide device level control (DLC), power management (PM) and energy management (EM) of MG systems \cite{2020_Bijaieh,2021_DSL}. The general form of this hierarchical framework enables integration of many local (on-site) control methods such as droop-based controls, maximum power point tracking (MPPT) and many more. It is important to understand that various control methods should be carefully matched and eventually benchmarked versus appropriate metrics that are defined for specific systems. For example, combination of droop control (as primary) and MPPT (as lower lever local) controls will require some sort of power curtailment due to the mismatch between variable generation and fixed power injection to the local point of common coupling (PCC). The interoperability of such combinations will also depend on strictness of the constraints that are defined for the electrical levels of the system. Control specifications, appropriate metrics, and guidelines for systematic control evaluations for various control levels are provided in IEEE 2030.7 and 2030.8 standards \cite{2017_IEEEstd2030_7, 2017_IEEEstd2030_8}. 

The holistic nature of MGs and their control adds complexity for the device level and system-level control design, analysis, evaluation and validation. From an investigator's perspective, the choice of MG models with appropriate fidelity is critical. There is always a trade-off between the fidelity of the chosen model and the required computation power to compile the model. The investigator must comprehend the critical assumptions, model constraints, the validity of the performed experiments as well as the appropriate platform for simulation.

There are numerous works dedicated to modeling of MG systems. While works such as \cite{2014_Shafiee, 2017_Fran} present modeling, control, and stability of DC MG systems, they do not provide systematic guidelines for mathematical coupling, as well as a systematic approach for control design of the underlying mathematical sub-systems. Power-flow and energy transfer models of microgrid systems with a generalized and systematic control approach for the overall system is offered in \cite{2015_Hassell, 2020_Bijaieh} but do not provide guidelines for system-level mathematical coupling or in case the system needs to be scaled. For example, solving those problems in a symbolic framework would be very hard to manage. A solution for scalability for such problems can be found in symbolic mathematical modeling platforms that use Modelica \cite{Fritz_2011}, where, blocks of code can be packaged as subsystems coupled through the fundamental definition of across (voltage) and through (current) variables. 

In all cases, the investigator must choose appropriate tools that fit best to solve the problem at hand. There are advantages of acausal modeling such as done in Modelica over typical causal modeling that fits well with ordinary differential equation (ODE) solvers such that of used in Simulink \cite{2020_simulink}. Generally, multi-physics systems modeled in hybrid differential algebraic equations (DAEs), and models with discrete states are more suitable to be solved in modelling approach such as Modelica. On the other hand, causal modeling is more structured in a sense that the system is decomposed into a chain of causal interacting blocks \cite{2013_Dizqah}. The problem of solving DAEs in simulation environments such as Simulink is very common. Moreover, inappropriate programming can also lead into appearance of algebraic loops which may cause severe computation burden. Implications of existence of algebraic loops in a model includes: inability of code generation for the model, the Simulink algebraic loop solver might fail to solve, and while the Simulink algebraic loop solver is active, the simulation may run slowly \cite{1999_Shampine, 1970_rabi, 1980_more}. Common solutions to the algebraic loop problem are: to introduce unit delays to the blocks, and, to turn the DAEs into ODEs by introducing additional states to aid the solver for easier solve. Adding additional delays to the model may lead to inaccurate solutions and introducing additional states might unnecessarily increase the size of the model. Hence, providing systematic approaches for mathematical modeling of such dynamical systems is a viable path to pursue.       

This work presents a mathematical modeling approach for MG systems. The aim is to define specific generating, transmitting and loading mathematically expressed subsystems in a modular way such that the designer would be still capable of scaling or extending the overall simulated system. While the work presented here might seem iterative in part with respect to previous efforts, the authors are compelled to use the result of this work due to its appropriateness for control system design. Moreover, based on the current pedagogy in fundamental controls in academia, students are pushed to utilize mathematical models rather than simulation packages such as  \cite{simscape} hence a modeling approach that aligns well with current pedagogy may be very useful. The characteristics of the mathematical modeling approach in this paper are: (1) Ease of modeling; where the investigator uses fundamental blocks to create ODEs. (2) Modular mathematical blocks connected according to a general connection convention. (3) A LBM which is fast to run and appropriate for control development specifically at EM level. (4) The model does not include algebraic loops and is readily-made for code generation, and subsequently real-time implementation in platforms such as Simulink Real-time. (5) The model can be run whole, or in part, on any platform that allows advanced mathematical operations. (6) The system may run at variable sub-system fidelity; reducing computation for certain applications. (7) Model and control development process is more aligned with common practices and pedagogy.   

The shortcomings associated with this type of modeling includes: (1) all the limitations and assumptions that are inherited from use of ideal, simplified, and average models. Integration of switching functionality in a MG setting is possible but significantly increases the complexity of the model. (2) The models are not suitable for fault analysis since the bandwidth (BW) of operation will not be typically sufficient. (3) More complexity due to management of voltage and current signals and dealing with both KVL and KCL separately at device and system levels.                    

This paper is organized as follows. Section II demonstrates the DC MG configuration and the corresponding sub-systems. Section III presents an example of the defined models as a generalized Shipboard Power System (SPS). The control development is shown in Section IV and in Section V, the overall model under the presented control is implemented in Simulink Real-time.  

%% file: system_model.tex
\begin{figure}[t]
	\centering
	\includegraphics[width=0.44\textwidth]{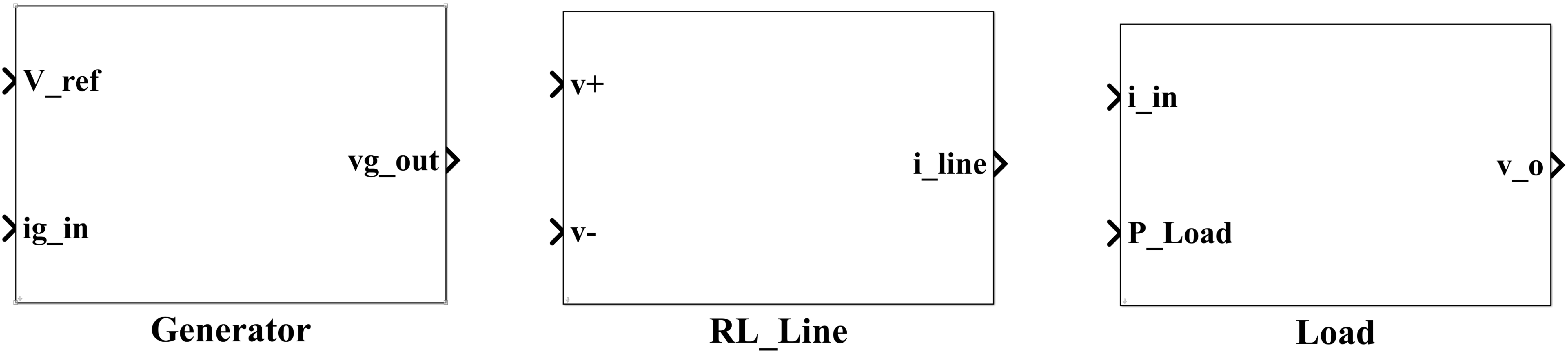}
	\caption{The overall generation, line, and load block configuration.   
    }
	\label{GenLineLoad}
\end{figure}
\begin{figure*}[t]
	\centering
	\includegraphics[width=0.75\textwidth]{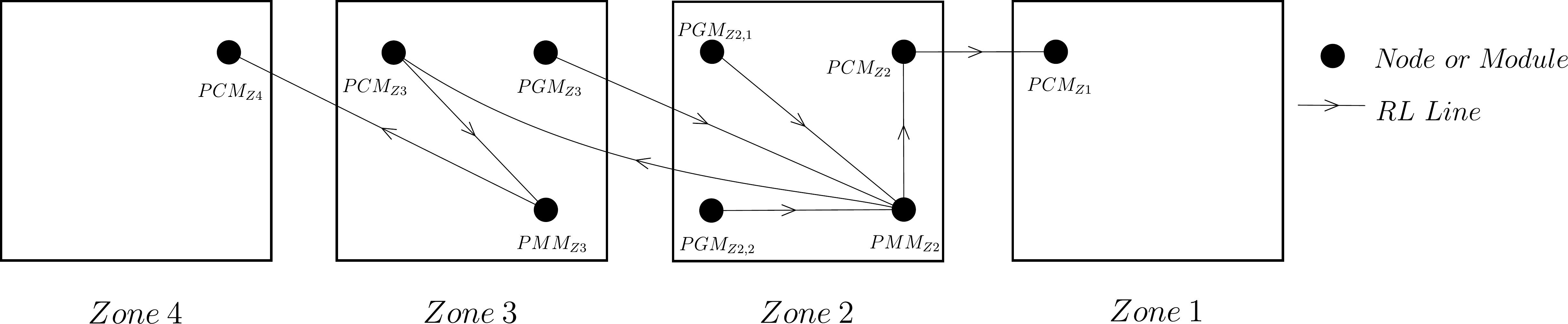}
	\caption{Connection convention using directed graph to specify KVL and KCL in the networked power system. 
    }
	\label{Graph}
\end{figure*}

DC MG systems consist of generation units and energy storage systems (ESSs) which supply a variety of loads through a common DC bus or through local transmission lines. Fig. \ref{GenLineLoad} demonstrates input/output configuration for generation, line and load modules. Generally the generator itself can contain ESSs as well as the local converters, rectifiers and loads. The line module is chosen in a specific form that couples the generation unit to the load module. The load module can contain point-of-load ESSs and various loads. In case a local bus exists, it may be coupled as a generation as well as a load module. One important aspect is that each generator and load module can only be connected through a line module which enables neighbor-to-neighbor and modular connection. Therefore, the overall system can then have an appropriate directed graph representation. Fig. \ref{Graph} shows a directed graph where, the nodes represent generation and load modules, and the edges show the coupling lines. The example Fig. \ref{Graph} includes a multi-zonal system that includes power generation, conversion, and load modules (PGMs, PCMs, and PMMs). In case there are multiple edges connected to a node, the corresponding local current injected to the load or generator nodes are obtained from KCL at the specific node hence enabling chain as well as loop connections. The overall microgrid model with the corresponding connection convention can then enable programmatical generation of the model.

%% file: dc_MG_model.tex
In this section, a generalized model of an SPS is presented considering the generator, line and load block input/output configurations. First, the overall multi-zone SPS is shown, then the corresponding models are presented.

\subsection{Notional Zonal Shipboard Power System}
\begin{figure}[t]
	\centering
	\includegraphics[width=0.49\textwidth]{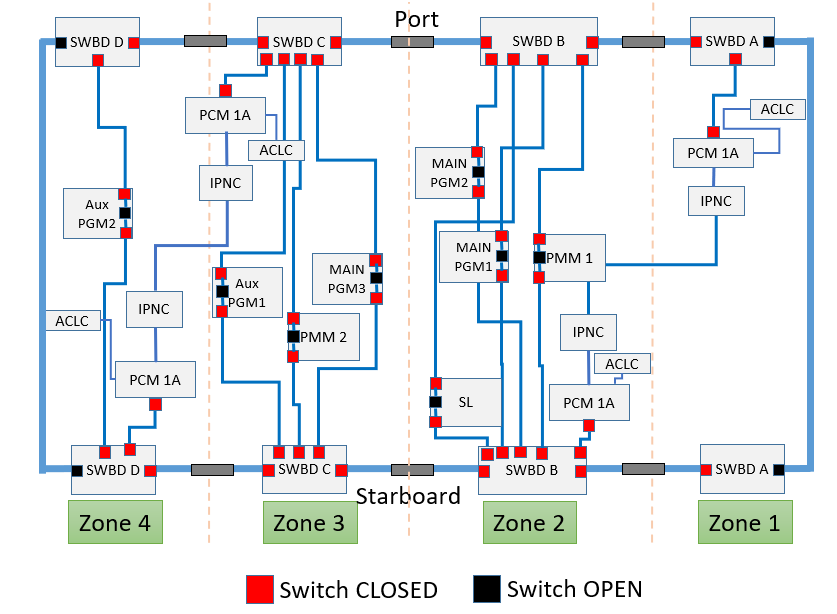}
	\caption{MVDC notional 4-zone ship DC microgrid system \cite{esrdc1270}.
    }
	\label{MVDC}
\end{figure}

In \cite{esrdc1270}, the notional SPS is divided into 4-zones where each include several subsystems which are mainly PGMs, PCMs and propulsion motor modules (PMMs). PGMs each include fuel operated generators, three-phase rectifiers, with the corresponding filters and control. The PCMs are defined to include ESSs, AC and DC loads and the underlying converters. All zones are inter-connected through various switches by a main $12~KV$ rated DC bus shown in Fig. \ref{MVDC}. In this work, for the sake of simplicity, among various modules in Fig. \ref{MVDC}, PGM, PCM and PMM are chosen as main system modules.

\subsection{Component Model Mathematical Representation}
\subsubsection{PGM}
\begin{figure}[t]
	\centering
	\includegraphics[width=0.47\textwidth]{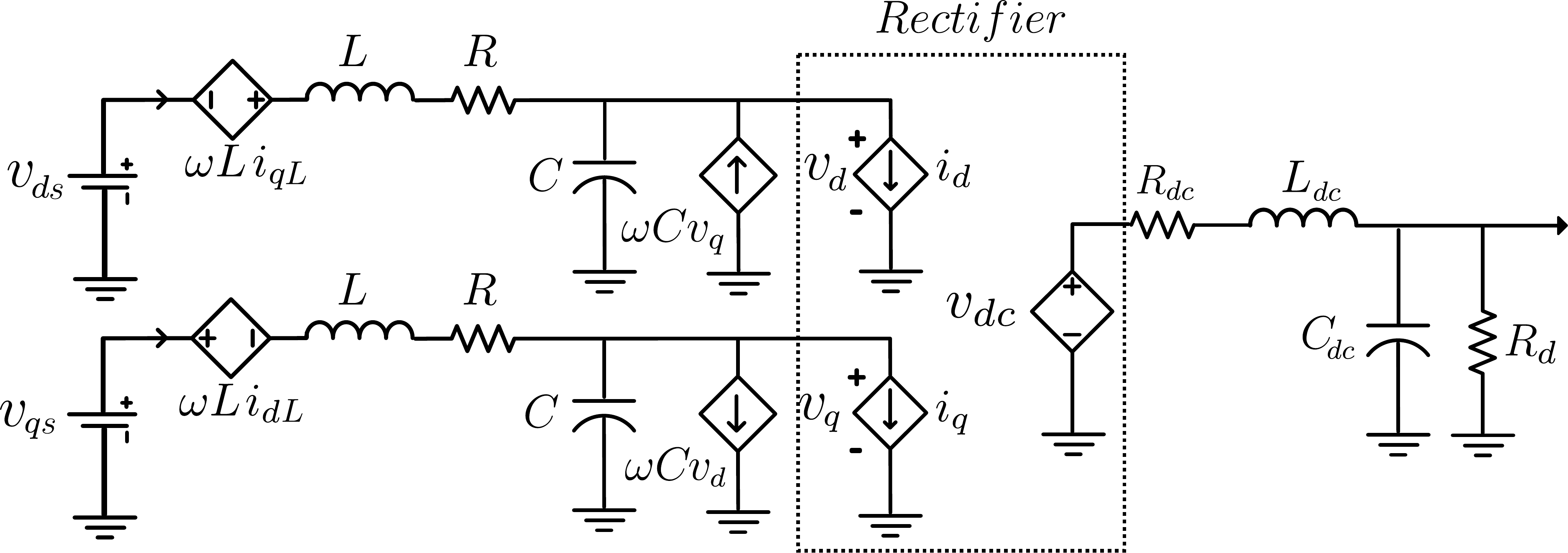}
	\caption{Power generation module including the AC-side, a rectifier and DC-side 
    }
	\label{PGM}
\end{figure}

Assuming the three-phase side of the system is balanced and the frequency of operation is fixed. Then the resulting rectifier model shown in Fig. (\ref{PGM}) is represented as follows: 
\begin{subequations} \label{rect_eq}
\begin{align}
    v_{dc} &= \dfrac{3}{2}(\lambda\dfrac{2\sqrt{3}}{\pi} v_d~ cos\phi+ \lambda\dfrac{2\sqrt{3}}{\pi}v_q~ sin\phi  ) \nonumber \\
          &=\dfrac{3 \sqrt{3}}{\pi} \lambda (v_d~ cos\phi +v_q~ sin\phi)\\
i_d &= \lambda \dfrac{2\sqrt{3}}{\pi} cos \phi~ i_{dc}   \\
i_q &= \lambda \dfrac{2\sqrt{3}}{\pi} sin \phi~ i_{dc}. 
\end{align}
\end{subequations}
Where, $\lambda=cos(\alpha)$, and $\alpha$ is the rectifier firing angle. The overall state-space representation of the PGM is:
\begin{subequations} \label{PGM_eq}%
\begin{align} 
   L\frac{di_{dL}}{dt} &= v_{ds} +\omega L i_{qL} - Ri_{dL} -v_d  \\
   L\frac{di_{qL}}{dt} &= v_{qs} -\omega L i_{dL} - Ri_{qL} -v_q  \\
   C\frac{dv_d}{dt} &= i_{dL} + \omega C v_q - i_d\\
   C\frac{dv_q}{dt} &= i_{qL} - \omega C v_d - i_q\\
   L_{dc}\frac{di_{dc}}{dt} &= v_{dc} - R_{dc} i_{dc} -v_{c, dc} \\
   C\frac{dv_{c,dc}}{dt} &= i_{dc} - \dfrac{v_{c,dc}}{R_d}-i_{g,in} 
\end{align}
\end{subequations}
It is important to note the PGM model in (\ref{PGM_eq}) is a simplified model that represents power flow and energy transfer in the system and might not provide adequate fidelity for many analysis approaches. For example, \cite{1996_Sudhoff} shows that the generation model itself can also depend on the rectifier control which adds more complexity to the design and analysis. However, the overall PGM module can be reformed to adhere to Fig. \ref{Graph}. The models here are chosen to be detailed enough to demonstrate the approach and the control objectives. Testing and verification of more detailed and higher fidelity models are out of the scope and are left for future iterations of this work.

\subsubsection{Line Module}

The overall input/output connection of the line module is shown in Fig. \ref{GenLineLoad}. The state space expression is defined as
\begin{equation} \label{line_mod}
    L_{line}\frac{di_{L,line}}{dt} = v_{in}-v_{out}-R_{line}i_{L,line},
\end{equation}
where, $v_{in}$ and $v_{out}$ are shown as $v_+$ and $v_-$ in Fig. \ref{GenLineLoad}. $L_{line}$ and $R_{line}$ are the RL inductance and resistance, and $i_{L,line}$ is the corresponding current. It is important to note that here $v_{in}$ and $v_{out}$ are the inputs, and $i_{L,line}$ is the output of the line module as shown in Fig. \ref{GenLineLoad}. Therefore, the state $i_{L,line}$ is computed after initialization and availability of $v_{in}$ and $v_{out}$. In simulation platforms such as Simulink, the state-space expression such as in (\ref{line_mod}) is solved using integrator blocks which dictate the output of the module.

\subsubsection{ESS}

The  ESS is considered to be ideal and may include single or hybrid storage systems such as battery energy storage systems (BESS), flywheels, or super-capacitor banks. Dynamics of the ESS can be expressed as:
\begin{align} \label{PCM_eqs}
\frac{{di}_{ESS}}{dt} &= \omega_{ESS}(i_{ESS,ref}-i_{ESS}),
\end{align}
where, $\omega_{ESS}$ represents the ESS response. $i_{ESS,ref}$ is the ESS reference signal and $i_{ESS}$ is the actual ESS injected current. Considering the bus voltage $v_b$, ESS reference power is $P_{ESS,ref}=v_bi_{ESS,ref}$ and the actual injected power is $P_{ESS}=v_bi_{ESS}$ \cite{Bijaieh_2020}. In case of a battery, the SOC can be calculated as:
\begin{align}\label{ESS_SOC}
SOC &= \frac{Q_{0}-\frac{1}{3600}\int{i_{batt}(t)}dt}{Q_T},
\end{align}
where, $Q_0$ and $Q_T$ are the the initial and total energy stored in the battery ESS in $AHr$, and $i_{batt}$ is the injected battery current. The SOC versus the injected power can be obtained as
\begin{align} \label{SOCp}
SOC= \dfrac{Q_0v_b-\dfrac{1}{3600} \int P_{batt}(t)dt}{Q_Tv_b},
\end{align}
 where, $v_b$ is the instantaneous measured voltage.

\subsubsection{Load Module}
The load is modelled as a controlled current source and a parallel $RC$ pair. The simplified mathematical expression of the load is defined as
\begin{equation} \label{load_eq}
    C_L\frac{dv_{c,L}}{dt} = -\frac{P_i(t)}{v_{c,L}} - \frac{v_{c,L}}{{R_L}}+i_{in},
\end{equation}
where, $v_{c,L}$ is the load bus voltage and $P_{i}(t)$ is the load demand, $R_L$ is the resistive load, and $i_{in}$ is the overall load module demand current injected corresponding to Fig. \ref{GenLineLoad}. Hereafter, a PMM includes a load module while a PCM includes a load and a shunt ESS module. In the next section, the power sharing control as well as the PGM and the ESS control is developed.

%% file: control_development.tex

\begin{figure}[t]
	\centering
	\includegraphics[width=0.44\textwidth]{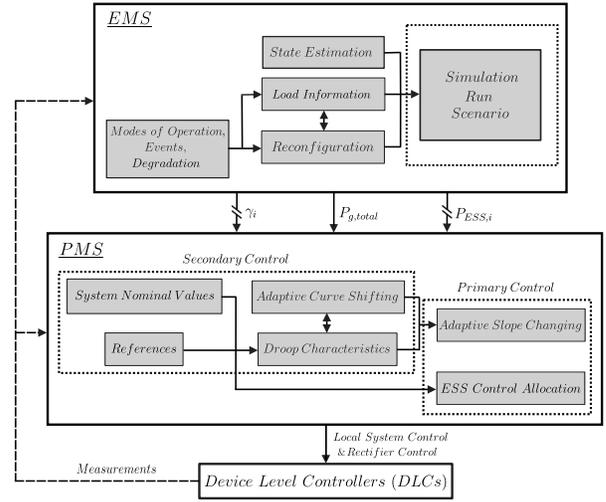}
	\caption{The control system hierarchy for SPS.  
    }
	\label{Control_Sys}
\end{figure}

\begin{figure}[t!]
	\centering
	\includegraphics[width=0.30\textwidth]{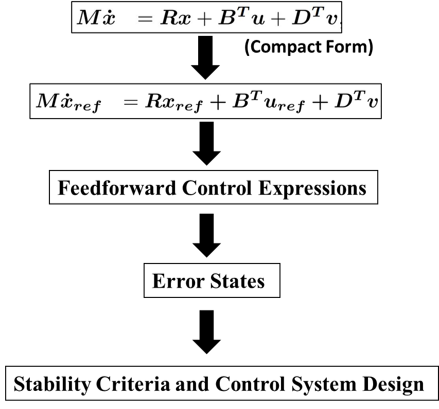}
	\caption{Control design process.  
    }
	\label{Control_Des}
\end{figure}

Fig. \ref{Control_Sys} shows the overall hierarchical control architecture. Considering modes of operation, degradation, and events (such as identification processes), and with subsequent reconfiguration as well as considering load information, a simulation run scenario is chosen for EM layer. The ESS control command and the generation power sharing weighting is fed into a PM layer where an adaptive droop control \cite{Mokhtar_2017} enforces the power sharing as well as maintaining the specified bus voltage. Fig. \ref{Control_Des} demonstrates the control design process \cite{2015_Weaver}. First, the state space systems is arranged in a compact form, then, the reference state-space system is defined. Next, the feedforward expressions are obtained from steady-state solution of the reference system. The amalgamation of the system with the feedforward control shifts the overall system to the origin where the error states can be analysed and used for stability and control system design. 

In this section the controls for the droop operation of the PGMs and the device level control for the underlying rectifiers and the ESSs are presented. 

\subsection{Primary Control: Adaptive Droop Control}
Considering the DC droop control, the $i$-indexed control commands using current measurements are
	\begin{align}
	v_{d,ref,i}&=v_{d,ref}-r_{d,i}i_{d,i}.
	\end{align} 	\label{droop_idq}%
$i_{d,i}$ is the injected bus current. $v_{d,ref,i}$, and $r_{d,i}$ are the droop characteristic settings. The adaptive droop \cite{Mokhtar_2017,2020_Bijaieh_fsu} is defined as 
\begin{align} \label{adaptive_droop}
v_{d,ref,i}&=v_{d,ref}-(r_{d,i,init}+\delta R_{d,i})i_{d,i}+\Delta v_{b}, 
\end{align}  
where, $v_{d,ref,i}$ is the droop control voltage commands, $v_{d,ref}$ is the initial droop voltage command, $r_{d,init,i}$ is the initial guess for virtual resistance value, $\delta R_{d,i}$ denote the droop resistance changes, and $\Delta v_{b}$ is the voltage deviation commands obtained from the secondary control loop and is obtained as
\begin{subequations} %
\begin{align}
\Delta v_{d}&=K_{p,v}(t)e_{v}(t)+K_{i,v}(t)\int_{0}^{\tau} e_{v}(\tau)d\tau\\
e_{v}&=v^*_{b}-v_{b},
\end{align} 
\end{subequations}
where, $e_v$ is the $12KV$ main bus error voltage. It is important to note for the MG system in the previous section, it is assumed that local lines are known, hence, $\delta R_{d,i}$ in (\ref{adaptive_droop}) is considered to be zero and power sharing is performed using curve shifting \cite{2020_Bijaieh_fsu}.


 \begin{table}[t!]
	\renewcommand{\arraystretch}{1.2}
	\caption{MG System and Control Parameters}
	\vspace{-1mm}
	\label{table_I}
	\centering
 \begin{tabular}{ | c | c | c| c | c | c|}
    \hline
     \multicolumn{6}{|c|}{\textbf{PGM}} \\ 
\hline
    \multicolumn{2}{|c|}{\textbf{Parameter}} & \multicolumn{2}{c|}{\textbf{Description}}  & \multicolumn{2}{|c|}{\textbf{Value}}
    \\ \hline
  \multicolumn{2}{|c|}{$L$} & \multicolumn{2}{c|}{3-ph System Inductance}  & \multicolumn{2}{|c|}{$100~\mu H$} 
    \\ \hline
    \multicolumn{2}{|c|}{$R$} & \multicolumn{2}{c|}{3-ph System Resistance}  & \multicolumn{2}{|c|}{$0.01~\Omega$} 
    \\ \hline
  \multicolumn{2}{|c|}{$C$} & \multicolumn{2}{c|}{3-ph System Capacitance}  & \multicolumn{2}{|c|}{$100~\mu F$} 
    \\ \hline
        \multicolumn{2}{|c|}{$f$} & \multicolumn{2}{c|}{3-ph System Frequency}  & \multicolumn{2}{|c|}{$120~Hz$} 
    \\ \hline
      \multicolumn{2}{|c|}{$L_{dc}$} & \multicolumn{2}{c|}{DC-side Inductance}  & \multicolumn{2}{|c|}{$200~\mu H$} 
    \\ \hline
    \multicolumn{2}{|c|}{$R_{dc}$} & \multicolumn{2}{c|}{DC-side Resistance}  & \multicolumn{2}{|c|}{$0.01~\Omega$} 
    \\ \hline
  \multicolumn{2}{|c|}{$C_{dc}$} & \multicolumn{2}{c|}{DC-side Capacitance}  & \multicolumn{2}{|c|}{$1~m F$} 
    \\ \hline
      \multicolumn{2}{|c|}{$R_{d}$} & \multicolumn{2}{c|}{Damping Resistor}  & \multicolumn{2}{|c|}{$10^{6}~\Omega$} 
    \\ \hline
     \multicolumn{6}{|c|}{\textbf{PCM}} \\ 
\hline
 \multicolumn{2}{|c|}{$Q_T$} & \multicolumn{2}{c|}{Battery ESS Capacity}  & \multicolumn{2}{|c|}{$10~AHr$} 
    \\ \hline 
 \multicolumn{2}{|c|}{$\omega_{ESS}$} & \multicolumn{2}{c|}{Battery ESS Response}  & \multicolumn{2}{|c|}{$1$} 
     \\ \hline 
 \multicolumn{2}{|c|}{$C_L$} & \multicolumn{2}{c|}{Load-side Capacitance}  & \multicolumn{2}{|c|}{$100~\mu F$} 
    \\ \hline 
         \multicolumn{6}{|c|}{\textbf{PMM}} \\ 
\hline
 \multicolumn{2}{|c|}{$C_{L,PMM}$} & \multicolumn{2}{c|}{PMM Capacitance}  & \multicolumn{2}{|c|}{$1~mF$} 
    \\ \hline 
    \end{tabular}

\vspace{0mm}
\end{table}    

\subsection{DLC Control: PGM Rectifier Control} 
The PGM feedforward and feedback control controls are obtained for (\ref{PGM_eq}). The feedforward control is obtained from the reference state-space model of (\ref{PGM_eq}), hence the rectifier feedforward  control is obtained from (\ref{rect_eq}a) and (\ref{PGM_eq}e) as
\begin{align}
\lambda&=\dfrac{\pi}{3\sqrt{3}}\dfrac{(R_{dc}i_{dc}+v^*_{c,dc})}{v_{d,ref} cos(\phi)+v_{q,ref} sin(\phi)}
\end{align} 
where, $v_{d,ref}$ and $v_{q,ref}$ are AC side reference values that are generally enforced through controlling $v_{ds}$ and $v_{qs}$. 

Hence, The overall feedback and feedforward control is  
\begin{subequations} 
\begin{align}
\lambda&=\dfrac{\pi}{3\sqrt{3}}\dfrac{(R_{dc}i_{dc}+v^*_{c,dc})}{v_{d,ref} cos(\phi)+v_{q,ref} sin(\phi)} \\
&+ K_p e(t)+K_{i}\int_{0}^{\tau} e(\tau)d\tau,
\end{align} 
\end{subequations} 
where, 
\begin{align}
e&=v^*_{c,dc}-v_{c,dc}.
\end{align}
Considering the hierarchical control architecture in Fig. \ref{Control_Sys}, $v^*_{c,dc}$ is fed from the droop control defined in (\ref{adaptive_droop}).

\subsection{ESS Control: Load Fluctuation Compensation} 
The ESS are controlled to compensate for high fluctuation and high ramp-rate portion of the loads of the specific zone. The overall control for the ESS systems are:
\begin{align} \label{PCM_eqs}
\frac{{dP}_{ESS,ref}}{dt} &= \omega_{ESS}(P_{Load}-P_{ESS,ref}),
\end{align}
where, $P_{Load}$ is the zone load and $P_{ESS,ref}$ is the reference that is fed into the corresponding ESSs. Next section presents a baseline example of the implementation of the control system.

%% file: simulation.tex
\begin{figure}[t!]
	\centering
	\includegraphics[width=0.44\textwidth]{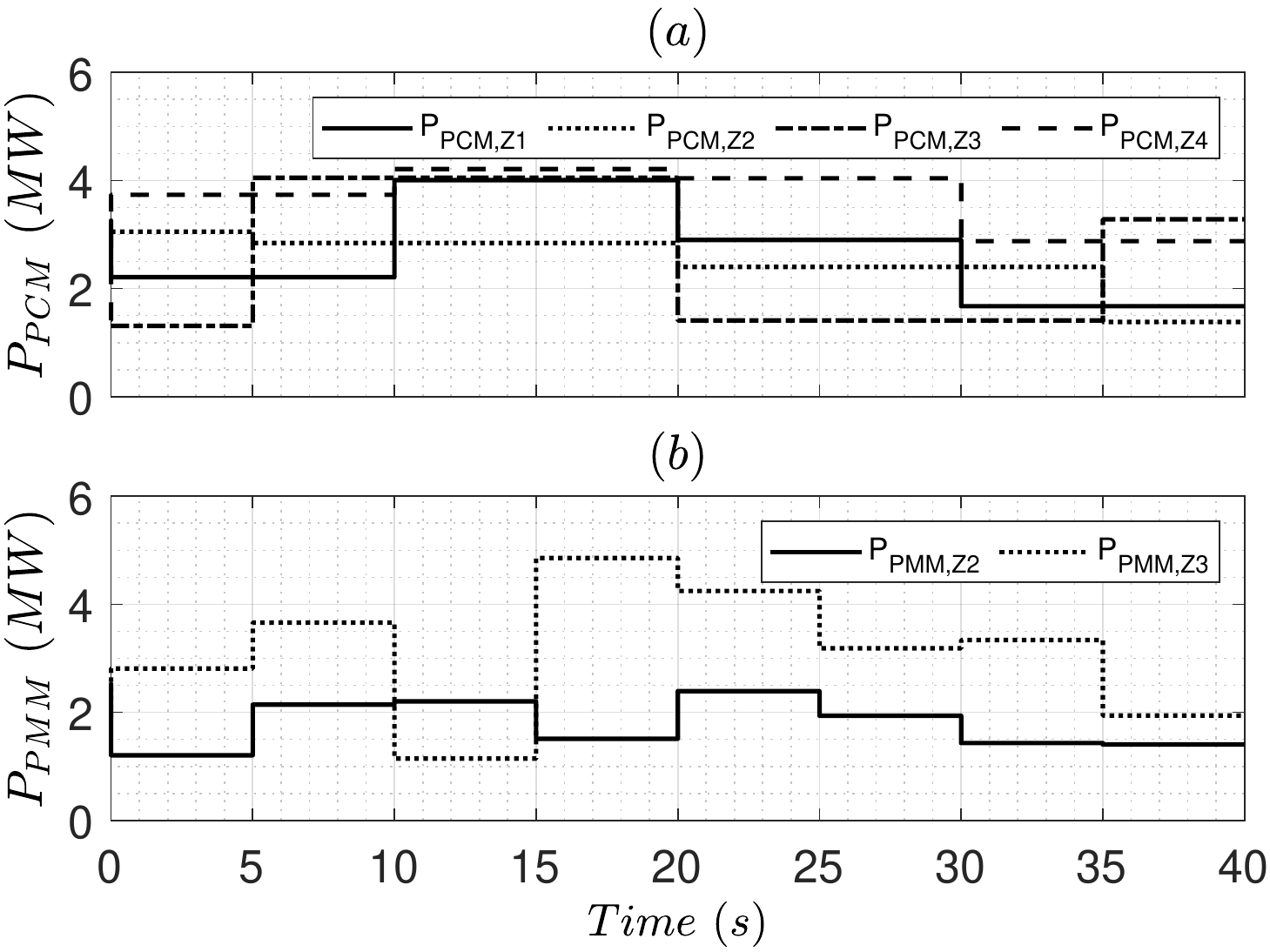}
	\caption{Individual PCM and PMM real-time loads 
    }
	\label{PLoad}
\end{figure}    

\begin{figure}[t!]
	\centering
	\includegraphics[width=0.44\textwidth]{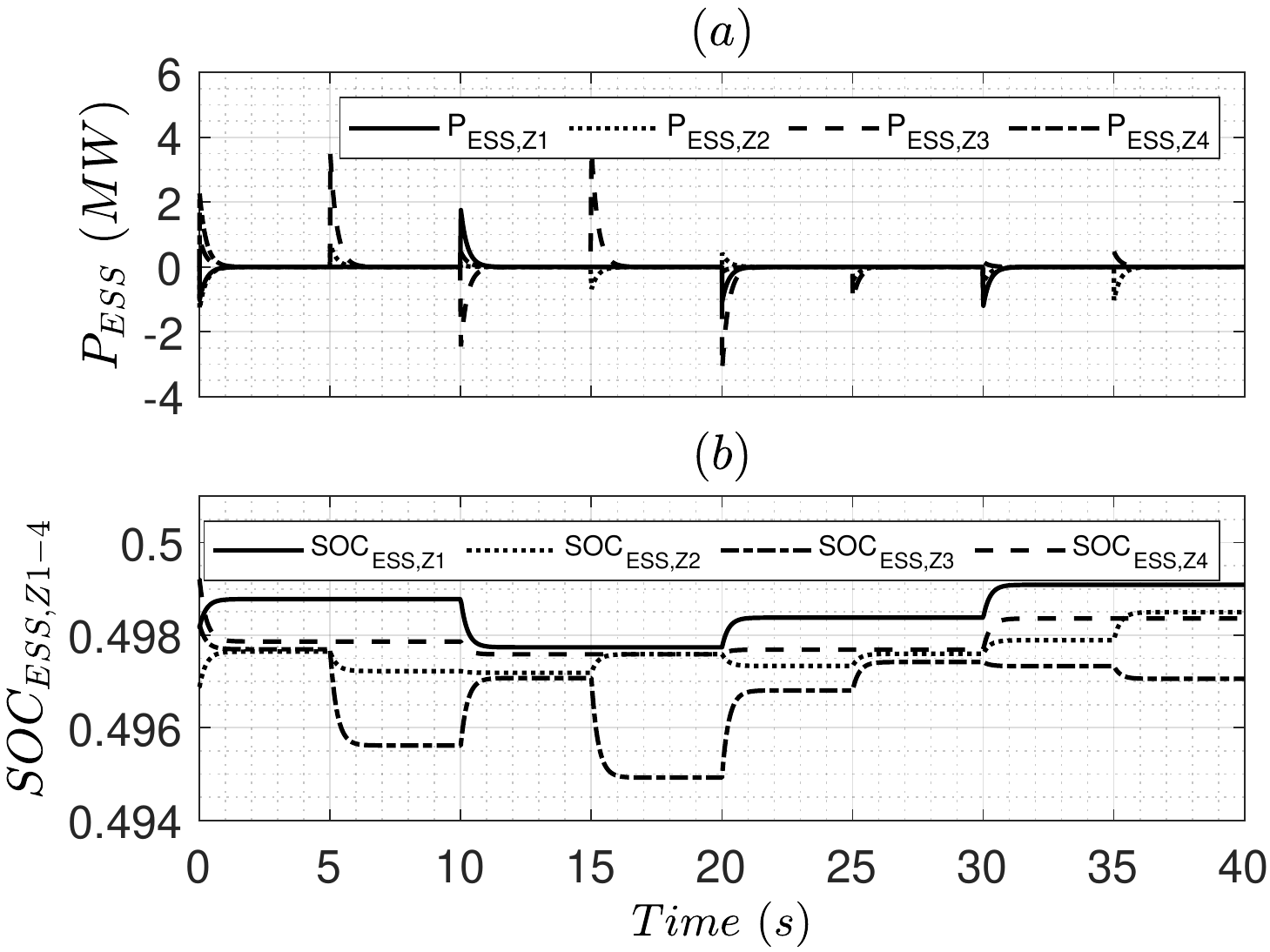}
	\caption{Energy storage systems, (a) instantaneous injected power and, (b) real-time processed energy through.  
    }
	\label{ESS_P_SOC}
\end{figure} 

\begin{figure}[t!]
	\centering
	\includegraphics[width=0.44\textwidth]{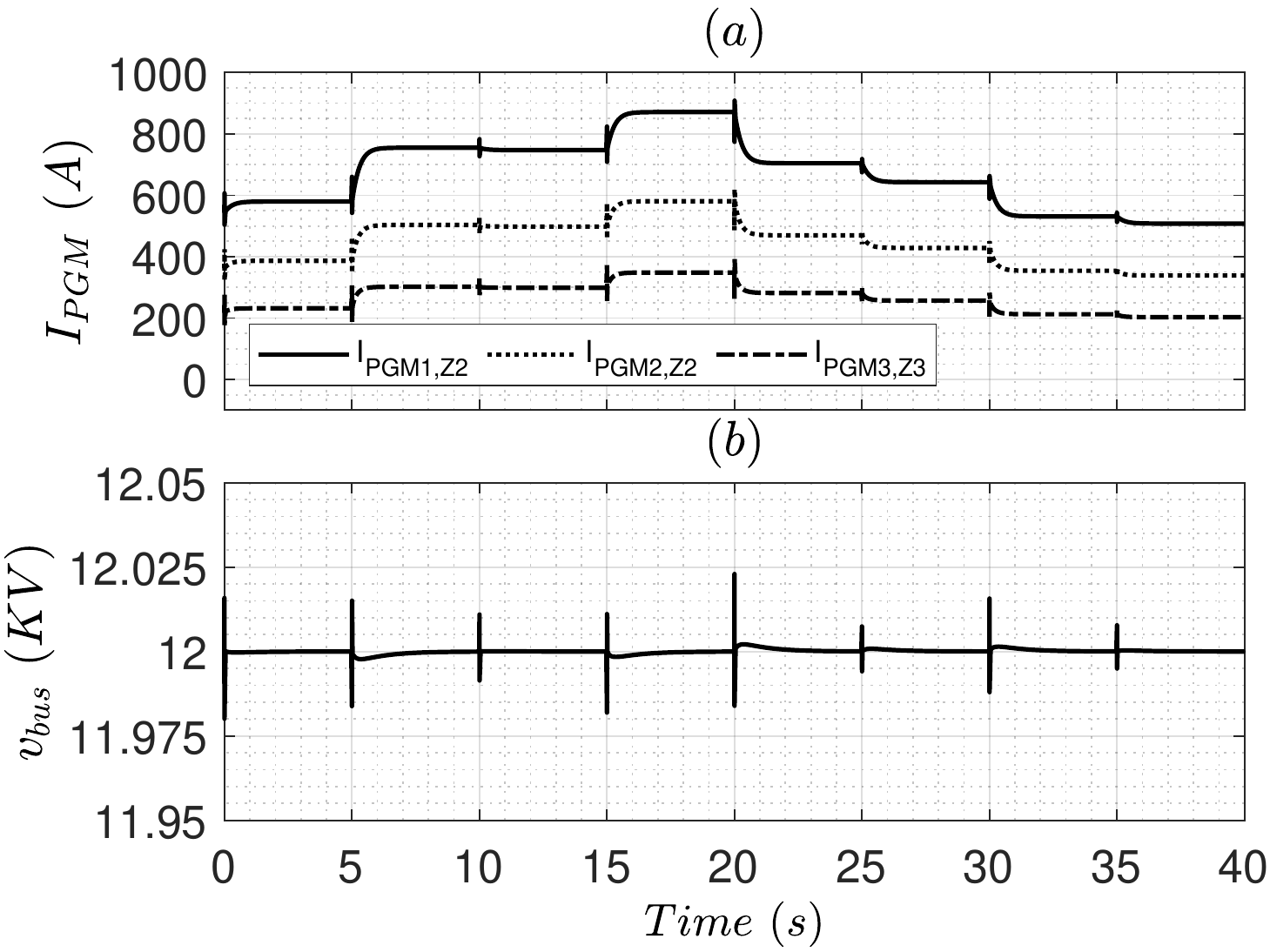}
	\caption{ (a) PGM power sharing and, (b) maintained 12 KV bus voltage.   
    }
	\label{igvb}
\end{figure} 

In this section the MG system in Section III, and the control in Section IV, with parameters shown in Table \ref{table_I} are used for a real-time simulation scenario. The aim of the test is to demonstrate the effectiveness of the model and the underlying control. The scenario includes demonstrating the behaviour of the PGMs under droop control while utilizing the ESSs for compensation of high fluctuation portion of the loads. The overall control diagram is shown in Fig. \ref{Control_Sys}. The results are shown in Fig. \ref{PLoad} to Fig. \ref{igvb}. 

Fig. \ref{PLoad} shows individual loads for PCMs and PMMs in the 4-zone SPS. The overall loads are stepped every $5s$ to demonstrate the behaviour of control against load step-changes. In this scenario the system is left to reach steady state before any future events are triggered.  

Fig. \ref{ESS_P_SOC} shows the ESS injected power and individual ESS SOCs. In Fig. \ref{ESS_P_SOC}a it can be seen that the ESSs effectively inject power when the systems faces high load fluctuations (high ramp-rate). Fig. \ref{ESS_P_SOC}b  shows the processed energy through the simulation time. It can be seen that ESSs charge and discharge where appropriate to meet the control requirements. 

Fig. \ref{igvb}a shows the injected currents of PGMs under the droop control. The droop is set so that there is a weighting of $5:3:2$. It can be seen that the power sharing is maintained through the simulation time. Fig. \ref{igvb}b show the main bus voltage of $12KV$. It can be seen that the main SPS voltage is  as expected since the ESSs effectively remove the high ramp-rate portion of the overall load.

 




%% file: conclusion.tex
This paper presented a simplified modular mathematical modelling approach for DC microgrid systems. It was shown that the overall mathematical model of the system can be directly implemented and networked through a specific connection convention. In this work, first a simplified mathematical model of a medium voltage DC (MVDC) shipboard MG system was presented. Then, the overall connection convention was shown. Next, a simplified example of the control development was presented, and the overall system under a baseline scenario was implemented in Simulink real-time.

%% file: Acknowledgement.tex
This material is based upon research supported by, or in part by, the U.S. Office of Naval Research (ONR) under award number N00014-16-1-2956.